\documentclass[prb,aps,twocolumn]{revtex4}
\usepackage{epsfig}
\newcommand{\be}{\begin{equation}}
\newcommand{\ee}{\end{equation}}
\newcommand{\bea}{\begin{eqnarray}}
\newcommand{\eea}{\end{eqnarray}}

\newcommand{\la}{\langle}
\newcommand{\ra}{\rangle}

\newcommand{\p}{\partial}

\def\nn{\nonumber\\}
\def\r#1{(\ref{#1})}

\begin{document}

\title{Dynamical density correlation function of 1D Mott insulators in
a magnetic field}

\author{Davide Controzzi$^{(a,b)}$ and Fabian H.L. Essler$^{(c)}$}
\affiliation{
$^{(a)}$ Department of Physics, Princeton University, Princeton NJ08544, USA\\
$^{(b)}$ International School for Advanced Studies, Trieste 34014, Italy\\
$^{(c)}$ Department of  Physics, Brookhaven National Laboratory, Upton, NY
11973-5000, USA
}
\begin{abstract}
We consider the one dimensional (1D) extended Hubbard model at half
filling in the presence of a magnetic field. Using field theory
techniques we calculate the dynamical density-density correlation
function $\chi_{nn}(\omega,q)$ in the low-energy limit. When excitons
are formed, a singularity appears in $\chi_{nn}(\omega,q)$ at a
particular energy and momentum transfer.
\end{abstract}

\maketitle

\section{Introduction}
Quasi 1D Mott insulators \cite{Mott} display unusual phenomena like
spin-charge separation and dynamical generation of a spectral gap and
have therefore attracted much attention in recent years. At present
the best realizations of 1D Mott insulators are found in anisotropic
antiferromagnets like ${\rm SrCuO_2}$ or ${\rm Sr_2CuO_3}$. The
dynamics of the latter compound has been 
studied both by angular-resolved photoemission \cite{arpes} and by
electron energy-loss spectroscopy (EELS) \cite{eels}. EELS measures the
dynamical density-density correlation function.
Theoretical descriptions of 1D itinerant antiferromagnets are based on
the half-filled Hubbard model with a large on-site repulsion $U$
(compared to the hopping matrix element $t$), which ensures that the
single-particle Mott gap is large. In the case of ${\rm
Sr_2CuO_3}$ it is believed that $U\approx 8t$ and that density-density
interactions between neighboring sites need to be taken into account
\cite{arpes,eels}.
Given that $U$ is ``much'' larger than $t$, strong-coupling expansions
around $U=\infty$ are an appropriate starting point. They have been
used to determine dynamical correlation functions \cite{Marburger,sp}
and to successfully model the EELS data for ${\rm SrCuO_2}$ \cite{eels}. 

There exist other materials believed to be quasi-1D Mott insulators
(e.g. the Bechgaard salts \cite{bech}), in which the Mott gap is small
compared to $t$.
The ``weak-coupling'' regime $U\alt 2t$ in which the Mott gap becomes
small is manifestly beyond the range of applicability of strong-coupling
expansions. The very existence of a gap  precludes the application of
conformal field theory \cite{FK,EF}.
However, this regime is accessible by an approach based on
exact field theory methods \cite{FF,lukyanov}. In Refs
\onlinecite{JGE,cet,EGJ,ET} this method has been used to determine the
optical conductivity and the single-particle Green's function for 1D
Mott insulators in the weak-coupling regime. Here we employ this
approach to calculate the dynamical density-density correlation function.
Motivated by suggestions that a magnetic field may generate X-ray
edge like threshold singularities \cite{kawakami} in the
density-density response, we take into consideration the effects of a
magnetic field. 
The outline of this paper is as follows: in section \ref{sec:FT} we
present the field theory description for the low energy degrees of
freedom. In section \ref{sec:corr} we determine the dynamical
density-density correlation function $\chi_{nn}(\omega,q)$. In section
\ref{sec:excitons} we consider an extended Hubbard model with
sufficiently strong nearest and next-nearest neighbor density-density
interactions. Here excitons are formed and we determine their
contributions to $\chi_{nn}(\omega,q)$. We summarize our results in
section \ref{sec:summ}.

\section{Field Theory description}
\label{sec:FT}

The extended Hubbard model in a magnetic field is described by the
Hamiltonian 
\bea
H &=& - t  \sum_{l;\sigma} 
\left( c_{l,\sigma}^\dagger c_{l+1,\sigma} + {\rm h.c.}\right) + 
U \sum_{l}  {n}_{l,\uparrow}{n}_{l,\downarrow}\nn
&&+\sum_{j=1}^2V_j\sum_{l}n_l n_{l+j}-\frac{h}{2}\sum_{l}\left (
{n}_{l,\uparrow} - {n}_{l,\downarrow} \right ),
\label{Hamiltonian}
\eea
where $c_{l,\sigma}$ are fermionic annihilation operators of spin
$\uparrow$ ($\sigma=1$) and $\downarrow$ ($\sigma=-1$),
$n_{l,\sigma}=c^\dagger_{l,\sigma}c_{l,\sigma}$ and
$n_l=n_{l,\uparrow}+n_{l,\downarrow}$.  For $h=0$ the ground state has
spin projection $S^z=0$ whereas for very large fields $h>h_c$ it is fully
polarized. We constrain our analysis to the case where the ground
state is partially magnetized, which corresponds to fields $0\leq h<h_c$.
A description of the low-energy degrees of
freedom of \r{Hamiltonian} for weak interactions $0<2V_2<2V_1<U\ll t$ is
then obtained by standard techniques \cite{GNT}. In the presence of
the field $h$ there are four Fermi points $\pm k_{F,\sigma}$ with
$k_{F,\downarrow}+k_{F,\uparrow}=\pi/a_0$, where $a_0$ is the lattice spacing. 
Taking into account only modes in the vicinity of $k_{F,\sigma}$ we expand  
\be
c_{l,\sigma}\longrightarrow \sqrt{a_0} \left[e^{ik_{F,\sigma} x}
\ R_\sigma(x)+e^{-ik_{F,\sigma} x}\ L_\sigma(x)\right],
\label{cpsi}
\end{equation}
where $x= l a_0$. The resulting
fermionic field theory can be bosonized with the result
\bea
{\cal L}_s &=&
\frac{1}{16\pi}\left[v_s(\p_x\Phi_s)^2-\frac{1}{v_s}(\p_{t}\Phi_s)^2
\right], 
\label{lspin}\\
{\cal L}_c &=&
\frac{1}{16\pi}\left[v_c(\p_x\Phi_c)^2-\frac{1}{v_c}(\p_{t}\Phi_c)^2 
\right]-\lambda\cos(\beta_c\Phi_c).\
\label{lagr}
\eea
The spin sector is a free bosonic theory whereas the charge sector is
described by the integrable Sine-Gordon model (SGM) \cite{SGM}.
Fermionic operators are expressed in terms of the canonical charge and
spin bose fields $\Phi_{c,s}$ and their respective dual fields
\be
\Theta_{c,s}(t,x)=\frac{-1}{v_{c,s}}\int_{-\infty}^xdy\
\partial_t\Phi_{c,s}(t,y)
\ee
by
\bea
L_{\sigma} &=&\eta_{\sigma}\ 
e^{\frac{i}{4}\left({\beta_c}\Phi_c-\frac{1}{\beta_c}\Theta_c\right)}\ 
e^{\frac{i}{4}{\sigma}({\beta_s}\Phi_s-\frac{1}{\beta_s}\Theta_s)},\nn
R_{\sigma} &=&\eta_{\sigma}\
e^{-\frac{i}{4}\left({\beta_c}\Phi_c+\frac{1}{\beta_c}\Theta_c\right)}\ 
e^{-\frac{i}{4}{\sigma}({\beta_s}\Phi_s+\frac{1}{\beta_s}\Theta_s)}.
\label{RL}
\eea
Here $\eta_{\sigma}=\eta_\sigma^\dagger$ are Klein factors that fulfill
$\lbrace\eta_\sigma,\eta_\tau\rbrace=2\delta_{\sigma,\tau}$. 
The spin and charge velocities $v_{c,s}$ and the parameters
$\beta_{c,s}$ depend on $t$, $h$, $U$ and $V$. 

\subsection{Hubbard model ($V_{1,2}=0$)}

For the Hubbard model $\beta_{c,s}$ and $v_{c,s}$  can be calculated
exactly from the Bethe Ansatz solution. The ``$\eta$-pairing'' SU(2)
symmetry of the half-filled Hubbard model \cite{so4} fixes
$\beta_c=1$, whereas $\beta_s(h,U)$ is obtained from the solution of a
linear integral equation \cite{woynarFSC,FK} 
\bea
\beta_s&=&{\sqrt{2} Z(\Lambda)}\ ,\nn
Z(\lambda)&=&1+\int_{-\Lambda}^{\Lambda}
d\mu\ a_2(\lambda-\mu)\ Z(\mu)\ ,
\eea
where $2\pi a_2(x)=U [x^2+U^2/4]^{-1}$. The integration boundary
$\Lambda$ is determined by the condition $\epsilon(\Lambda)=0$, where 
\bea
\epsilon(\lambda)&=&h-4{\rm Re}\sqrt{1-(\lambda-iU/4)^2}+U\nn
&&+\int_{-\Lambda}^{\Lambda} d\mu\ a_2(\lambda-\mu)\ \epsilon(\mu)\ .
\label{eps}
\eea
\begin{figure}[ht]
\begin{center}
\epsfxsize=0.45\textwidth
\epsfbox{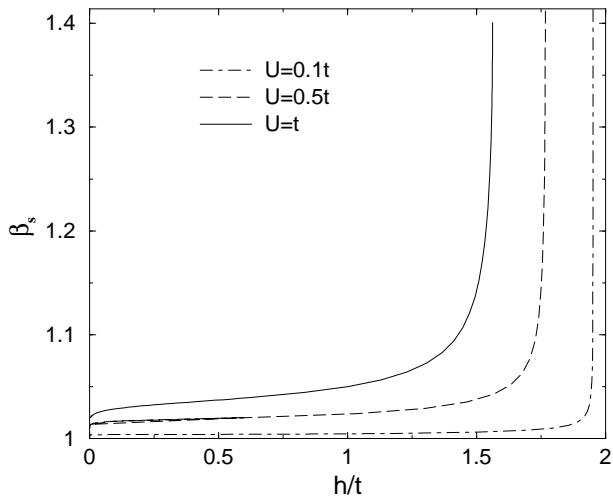}
\end{center}
\caption{\label{fig:z}
$\beta_s$ for the Hubbard model as a function of $h$ for different
values of $U$. 
}
\end{figure}
In the limit ${U}/{h}\to 0$ $\beta_s$ can be calculated
analytically by means of a Wiener-Hopf analysis
\be
\beta_s\simeq 1+\frac{U}{16\pi t\cos(\pi{\cal M})}\ ,
\ee
where ${\cal M}$ is the magnetization that is calculated from
the solution to an integral equation similar to \r{eps} (see
Ref. \onlinecite{FK}). 
We note that $\beta_s$ varies between $1$ for $h=0$ and
$\sqrt{2}$ for $h\to h_c$. The behavior of $\beta_s$ as a function
of $h$ for different values of $U$ is shown in Fig.\ref{fig:z}.
We see that for small values of $U$ $\beta_s$ remains very close to
$1$ up to large fields very close to $h_c$.

The spin and charge velocities can be calculated in a similar way.
Field theory is exact in the {\sl scaling limit}, which has been
constructed in the absence of a magnetic field in Ref.\onlinecite{ezer}
using exact results for the half-filled Hubbard model \cite{smat}.
The $h>0$ case can be mapped onto the attractive Hubbard model below
half-filling by means of the particle-hole transformation for spin
down $c_{j,-1}\rightarrow (-1)^j
c^\dagger_{j,-1}$. The scaling limit for the latter model has
been found by Woynarovich and Forgacs \cite{wf} and is obtained by
taking $t\to\infty$, $U\to 0$ while keeping 
\bea
&& \sqrt{Ut\cos^3(\pi{\cal M})}e^{-\pi t\cos(\pi{\cal M})/2U}={\rm fixed}.
\eea
In this limit $\beta_s=1$ and the low-energy effective field theory is
$SU(2)\times SU(2)$ symmetric. However, on the level of the
large-distance asymptotics of correlation functions of the underlying
Hubbard model this enhanced symmetry is broken down to $SU(2)\times U(1)$
by the oscillating factors in \r{cpsi}. For example, the leading
asymptotical behavior of the spin-spin correlation functions is
\bea
\langle S^x(x) S^x(0)\rangle&=&\langle S^y(x) S^y(0)\rangle=
\frac{B(-1)^{x/a_0}}{x}+\ldots\ ,\nn
\langle S^z(x) S^z(0)\rangle&=&\frac{A\cos 2k_{F,\sigma}x}{x}+\ldots\ ,
\eea
and the spin-$SU(2)$ symmetry is broken.
Previous experience \cite{JGE,EGJ} suggests that field theory gives a
good description of the lattice model in an extended vicinity of the
scaling limit, provided the gap is small compared to $t$. This is the
case as long as $U \alt 2t$. We will apply field theory in the regime
defined by this criterion and therefore allow $\beta_s$ to be
different from $1$. 

\subsection{Extended Hubbard model}

For $V_{1,2}\neq 0$ no exact results for the lattice model
\r{Hamiltonian} are available. However, the low-energy degrees of freedom of
\r{Hamiltonian} in the regime $0<V_2<V_1<\frac{U}{2}\alt t$ are still
described by \r{lspin}-\r{lagr}, but now with $\beta_c<1$ \cite{vic}. Recently
it was shown in Ref.\onlinecite{EGJ} that for sufficiently large
values of $U,V_1,V_2$ it is possible to reach the attractive regime of
the SGM $\beta_c<1/{\sqrt{2}}$, in which excitonic holon-antiholon bound
states form. We will first consider the range
${1}/{\sqrt{2}}<\beta_c\leq 1$, where no excitons exist and field
theory results for the optical conductivity \cite{cet} have been found
to be in good agreement with dynamical density matrix renormalization
group computations for the lattice model \r{Hamiltonian} \cite{eric}.  
In section \ref{sec:excitons} we extend the analysis to the regime
$\beta_c<1/\sqrt{2}$.

\subsection{Density Operator}

The density operator is expressed in terms of the spin and charge
bosonic fields as \cite{GNT}
\bea
n (x,t)&=& n_{0}(x,t)+\sum_\sigma n_{2k_{F,\sigma}}(x,t)\ ,\nn
n_0(x,t)&=&A\ \partial_x\Phi_c\ ,\nn
n_{2k_{F,\sigma}}(x,t)&=&A^\prime\ e^{2ik_{F,\sigma}x}\ 
\sin(\frac{\beta_c}{2}\Phi_c)\
e^{\frac{i\sigma\beta_s}{2}\Phi_s}\ .
\label{dens.bos}
\eea
Here $A$ and $A^\prime$ are numerical constants. 
For a less than half-filled band there is an additional contribution
\cite{GNT} 
\begin{equation}
n_{U}(x,t)=A_{U}\ \cos\left(2[k_{F,\uparrow}+k_{F,\downarrow}]x+
\beta_c\Phi_c\right),
\label{nu}
\end{equation}
which is obtained by integrating out the high-energy degrees of
freedom in the path integral representation for the density-density
correlation function of the lattice model. The operator \r{nu}
corresponds to scattering processes involving two particles and two
holes with momentum transfer $2(k_{F,\uparrow}+k_{F,\downarrow})=
2\pi/a_0$. As a result $A_{U}$ is proportional to $U/t$ and thus is small.
At half-filling we have $A_U=0$. This can be established by
considering the following discrete symmetry of the lattice model
\begin{equation}
c_{j,\sigma}\longleftrightarrow (-1)^j c^\dagger_{j,-\sigma}\ .
\end{equation}
The lattice density operator transforms as $n_j\to 2-n_j$.
In the field theory this symmetry corresponds to inverting the signs
of the charge boson and its dual field $\Phi_c\to -\Phi_c$,
$\Theta_c\to -\Theta_c$. The (normal ordered) density operator
$n(x,t)$ must transform to $-n(x,t)$ under this change of sign and
this implies that $A_U=0$. 

\section{Density Correlations for $\frac{1}{2}<\beta_c^2\leq 1$}
\label{sec:corr}

The density-density correlation function is given by
\bea
\label{gnn}
&&G_{nn}(x,t)=
G_{nn}^{0}+\sum_\sigma G_{nn}^{2k_{F,\sigma}}\ ,
\eea
where
\bea
G_{nn}^{0}&=&\la 0| n_0(x,t)\ n_0(0,0)|0\ra\ ,\nn
G_{nn}^{2k_{F,\sigma}}&=&\la 0| n_{2k_{F,\sigma}}(x,t)
\ n_{2k_{F,-\sigma}}(0,0)|0\ra\ .
\label{gnn2}
\eea
We note that there are no ``mixed terms'' as can be shown by
exploiting the transformation properties under charge conjugation.
Due to spin-charge separation the correlation functions in \r{gnn2}
factorize into spin and charge pieces. In the spin sector \r{lspin} we
are dealing with a simple Gaussian model and elementary considerations
give
\bea
_s\la 0 | e^{i\frac{\beta_s}{2}\Phi_s(x,t)}\
e^{-i\frac{\beta_s}{2}\Phi_s(0)}|0 \ra_s
=\left[ x^2 - (v_s t+i\epsilon)^2   \right ]^{-d},\nn
\label{spin}
\eea
where $|0\rangle_s$ denotes the vacuum in the spin sector and
\be
d=\beta_s^2/2.
\ee 
In the Hubbard model the exponent $d$ varies between $\frac{1}{2}$ for
zero field and $1$ for $h\to h_c$. In order to determine the charge
part of the correlators \r{gnn2} we make use of the integrability of
the SGM \r{lagr} 
describing the charge sector. In the range of $\beta_c$ considered here
the spectrum of the SGM consists of scattering states of {\sl
solitons} and {\sl antisolitons}, which are particles
of mass $M$, charge $Q=\pm e$ and relativistic dispersion
$e(p)=\sqrt{p^2+M^2}$. In the Hubbard model they correspond to holons
and antiholons respectively. It is convenient to parametrize energy
and momentum in terms of the rapidity variable $\theta$
\begin{equation}
\label{ep} 
p=\frac{M}{v_c} \sinh\theta, \; e = M\cosh\theta.
\end{equation}
We introduce an index $\varepsilon = \pm$ for solitons and antisolitons.
Then a scattering state of $n$ solitons/antisolitons with rapidities
$\{ \theta_k\}$ and internal indices $\{\varepsilon_k\}$ is denoted by 
$|\theta_n \ldots \theta_1 \rangle_{\varepsilon_n,\ldots,\varepsilon_1}$.
In the spectral representation of this basis of (anti)soliton
scattering states we may express the two-point function of an operator 
${\cal O}$ in the charge sector as
\bea
\label{corr}
_c\langle 0|{\cal O}(x,t) {\cal O}^\dagger(0)|0 \rangle_c
=\sum_{n=0}^\infty\sum_{\varepsilon_i}\int
\frac{d\theta_1\ldots d\theta_n}{(2\pi)^nn!}\qquad\quad&&\nn
\times\exp\biggl[{i\sum_{j=1}^n e_j t-p_jx}\biggr]
|_c\langle 0 | {\cal O}(0)|\theta_n\ldots\theta_1
\rangle_{\varepsilon_n\ldots\varepsilon_1}|^2.&&
\eea
Here $p_j$ and $e_j$ are given by (\ref{ep}), and the form factors
$_c\langle 0 | {\cal O}(0)|\theta_n\ldots\theta_1
\rangle_{\varepsilon_n\ldots\varepsilon_1}|$ can be calculated by
exploiting the integrability of the SGM \cite{FF,lukyanov}. As a
consequence of the transformation properties under charge conjugation
of the operators appearing in (\ref{gnn}), only intermediate states
with an even number of particles will contribute to (\ref{corr}).
In order to obtain an accurate result for the large-distance
asymptotics it is sufficient to take into account intermediate states
with only a small number of particles in the spectral sum \r{corr} 
\cite{oldff1}. For the case at hand we have \cite{FF,lukyanov}
\bea
|_c\langle 0|\partial_x \phi_c|\theta_1,\theta_2 \rangle_{\pm\mp}|^2 &=&
|f(2\theta_{-})\sinh\theta_+|^2,\ \quad\\
|_c\langle 0 |\sin(\frac{\beta_c\Phi_c}{2})
|\theta_1,\theta_2\rangle_{\pm\mp}|^2 
&=&Z_1 |f(2\theta_{-})|^2,
\eea
where $\theta_\pm=(\theta_1\pm\theta_2)/2$, $Z_{1}$ is a known 
constant \cite{lukyanov} and  
\bea
f(\theta)&=&\frac{F(\theta)}{\cosh \left 
( \frac{\theta+i\pi}{2\xi}\right )}\ ,\nn
F(\theta)&=&\sinh\!{\frac{\theta}{2}} \exp\!\left[\!\int_0^\infty\!\!\frac{dk}{k}
\frac{\sin^2 (\frac{k}{2}[\frac{\theta}{\pi}+i])
\sinh (\frac{1-\xi}{2}k)}
{\sinh \left ( \frac{\xi k}{2} \right ) 
\cosh\left( \frac{k}{2} \right)
\sinh k} \right]\!\!,\nn
\xi&=&\frac{\beta_c^2}{1-\beta_c^2}\ .
\eea
EELS measures the imaginary part of the retarded dynamical
density-density correlation function
\bea
\chi_{nn}(\omega,k)&=& {\rm Im}\biggl\{i\int_{-\infty}^\infty dx
\int_{0}^\infty dt 
\ e^{i \omega t-ikx}\Bigl[G_{nn}(x,t)\nn
&&\qquad-G_{nn}(-x,-t)\Bigr]\biggr\}
\label{chinn}
\eea
We evaluate $\chi_{nn}(\omega,k)$ in the vicinity of the low-energy
modes at $k=0,2k_{F,\sigma}$ by Fourier transforming the
large-distance asymptotics of the density-density correlation
function. The latter is obtained by carrying out the form factor
expansion \r{corr} in the charge sector and multiplying it by the
spin-piece \r{spin} in the case of the $2k_{F,\sigma}$ response.

\subsection{Small $k$ behavior}
In the vicinity of $k=0$ the dynamical density response is dominated
by the contribution from $G^0_{nn}(x,t)$. This contribution does not
involve the spin sector and straightforward calculations give
\bea
\chi_{nn}(\omega,q)&=& A^2 8M^2
\frac{(v_c q)^2|f_0(\theta_0)|^2}{s^3 \sqrt {s^2-4M^2}}
\Theta(s^2-4M^2),\nn
\label{q0}
\eea
with $s^2(\omega,q)=\omega^2-(v_c q)^2$ and
$\theta_0=2$arcosh$(s/2M)$. Above the threshold at
$\omega=\sqrt{v_c^2q^2+4M^2}$, $\chi_{nn}(\omega,q)$
increases from zero in a universal square root fashion. This is due to
the momentum dependence of the form factors.

From the Heisenberg equations of motions for the lattice density
operator one can derive a relation between $\chi_{nn}$ and the optical
conductivity $\sigma(\omega)$ 
\begin{equation}
{\rm Re}\ \sigma(\omega)=\lim_{q\to
0}\frac{\omega}{q^2}\ \chi_{nn}(\omega,q)\ .
\label{eom}
\end{equation}
The optical conductivity has been calculated in \cite{cet} and agrees
with \r{eom} and \r{q0}. We note that a contribution of the type
\r{nu} to the density operator would violate the relation \r{eom},
which is an independent argument showing that $A_U= 0$
\cite{conteq}. 

\subsection{Behavior around $k=2k_{F,\sigma}$}

In the vicinity of $k=2k_{F,\sigma}$ the dynamical density response is
dominated by the contribution from $G^{2k_{F,\sigma}}_{nn}(x,t)$ and
involves both the spin and the charge sector. The threshold can be
determined by considering the lowest intermediate state that couples
to the density operator at $k=2k_{F,\sigma}$, which is a scattering
state of one soliton, one antisoliton and one spinon. The total
momentum and energy of this state are
\bea
P&=&p+q_1+q_2\ , \nn
E&=&v_s|p|+\sum_{j=1}^2\sqrt{M^2+v_c^2q_j^2}\ .
\eea
The threshold is obtained by minimizing the energy at fixed total
momentum with respect to $p,q_1,q_2$
\bea
E_{\rm thres}&=&\min_{q_1,q_2}\left[
v_s|P-q_1-q_2|+\sum_{j=1}^2\sqrt{M^2+v_c^2q_j^2}\right]\nn
&=&\cases{
\sqrt{4M^2+v_c^2 P^2} & {\rm if}$\ |P|\leq Q$\cr
v_s|P|+2M\sqrt{1-\alpha^2} & {\rm if}$\ |P|\geq Q$\cr
}\ ,
\label{thres}
\eea
where
\be
\alpha=\frac{v_s}{v_c}\ ,\quad
Q = \frac{2M v_s}{v_c\sqrt{v_c^2 - v_s^2}}\ .
\ee
The behavior is quite similar to what is found for the spectral
function \cite{ET}.

\subsubsection{Equal velocities $v_s=v_c=v$}

For $v_s=v_c=v$ one can obtain the following representation
\bea
&&\chi_{nn}(\omega,2k_{F,\sigma}+q)\approx
\frac{\Gamma^2(1-d)Z_1{A^\prime}^2}
{\pi(2v)^{2d-1}}\!\int_{-\infty}^\infty\!\!\!d\theta\frac{|f_1(2\theta)|^2}
{c(\theta)^{2-2d}}\nn
&&\times \ {\rm Im}F\left(1-d,1-d,1,\frac{\omega^2-v^2q^2}{c^2(\theta)}\right),
\label{equalv}
\eea
where $c(\theta)=2M\cosh(\theta)$. The imaginary part of the
hypergeometric function vanishes unless
$c(\theta)<\sqrt{\omega^2-v^2q^2}$. This implies that the response
function is nonzero only if $\omega^2>v^2q^2+4M^2$, in
agreement with \r{thres}.  
Just above the threshold ($0<s/2M\ll 1$) we may use the
transformation formulas for hypergeometric functions to obtain
($s^2=\omega^2-v^2q^2$)
\bea
&&{\rm Im}F\left(1-d,1-d,1,\frac{s^2}{c^2(\theta)}\right)
=-\frac{\Gamma(1-2d)\ \sin(\pi 2d)}{\Gamma^2(1-d)}\nn
&&\times
F\left(d,d,2d,1-\frac{s^2}{c^2(\theta)}\right)
\left(\frac{s^2}{c^2(\theta)}-1\right)^{2d-1}\nn
&&\times\ \Theta(s^2-c^2(\theta)),
\eea
where $\Theta(x)$ is the Heaviside function.
The remaining $\theta$-integral in \r{equalv} is therefore over a very
small interval $[-{\rm arccosh}(s/2M),{\rm arccosh}(s/2M)]$ and can be
taken by Taylor-expanding the integrand. The leading contribution
to the behavior just above the threshold is
\bea
\chi_{nn}(\omega,2k_{F,\sigma}+q)&\propto&
\left(\frac{s-2M}{M}\right)^{\frac{1}{2}+2d}\ ,\nn
&& \frac{s-2M}{M}\to 0.\qquad
\label{thresbehav}
\eea
Here we have used that $\beta_c^2>1/2$, in which case we have
\begin{equation}
f_1(2\theta)\propto \theta\ ,\ {\rm for}\ \theta\to 0.
\end{equation}
At the Luther-Emery point $\beta^2_c=\frac{1}{2}$ \cite{LuE} there is
a different power law increase in \r{thresbehav} (the exponent is
$2d-\frac{1}{2}$). As $\beta_c^2\to \frac{1}{2}$ from above the region
is which \r{thresbehav} holds shrinks to zero.
The important result here is that $\chi_{nn}$ vanishes as the
threshold is approached from above. There are no threshold
singularities! The behavior for large frequencies
$\omega\gg\sqrt{v^2q^2+4M^2}$ (but necessarily $\omega\ll t$ for field
theory to apply) is 
\begin{equation}
\chi_{nn}(\omega,2k_{F,\sigma}+q)\propto
s^{2d-2+\beta_c^2}\ ,\quad s\gg M.
\end{equation}

\subsubsection{Different velocities $v_s\neq v_c$}

In the case of different spin and charge velocities
$\frac{v_s}{v_c}\equiv\alpha < 1$ one may represent $\chi_{nn}$ as 
\bea
&&\chi_{nn}(\omega,2k_{F,\sigma}+q)\approx
\frac{Z_1{A^\prime}^2}{\Gamma^2(d)(2v_s)^{2d-1}}\nn
&&\times\int_{-\infty}^\infty d\theta_+ d\theta_-
|f_1(2\theta_-)|^2
\left(\Omega\Omega^\prime\right)^{d-1}
\Theta(\Omega)\ \Theta(\Omega^\prime),\nn
\label{differentv}
\eea
where
\bea
\Omega&=&\omega-v_sq-2M\cosh(\theta_-)
\left[\cosh(\theta_+)-\alpha\sinh(\theta_+)\right],\nn
\Omega^\prime&=&\omega+v_sq-2M\cosh(\theta_-)
\left[\cosh(\theta_+)+\alpha\sinh(\theta_+)\right].\nn
\eea
One easily checks that \r{differentv} leads to a threshold described
by \r{thres}. The remaining integrals in \r{differentv} are evaluated
numerically. 

\begin{figure}[ht]
\begin{center}
\epsfxsize=0.45\textwidth
\epsfbox{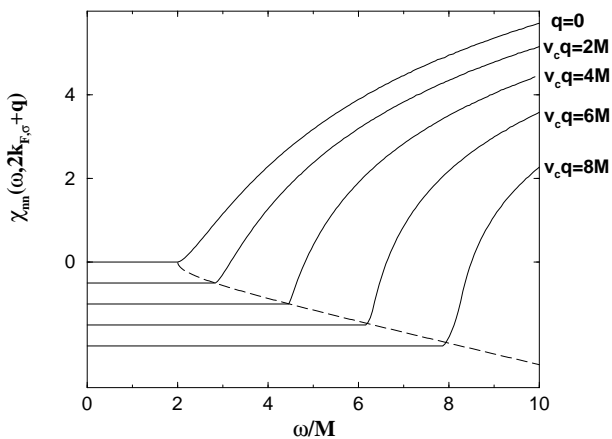}
\end{center}
\caption{\label{fig:chinn}
$\chi_{nn}(\omega,2k_{F,\sigma}+q)$ for $\beta_c=\beta_s=1$,
$\alpha=0.851$ and several different values of $v_cq$. The curves have
been offset. The dashed line is the threshold when spinons carry zero
momentum $\sqrt{4M^2+v_c^2q^2}$.
}
\end{figure}

A remarkable feature of the half-filled Mott insulator is the presence
of two dispersing features in the spectral function, that are
associated with $v_s$ and $v_c$ respectively \cite{arpes,ET}. A
natural question is whether an analogous feature exists in the
density-density response. We have analyzed \r{differentv} for several
sets of parameters $\beta_c$, $\beta_s$, $\alpha$ and found that in
all cases $\chi_{nn}(\omega,2k_{F,\sigma}+q)$ is rather
featureless. There are no singularities or peaks that can be
associated with $v_c$ and $v_s$ separately. We also do not find any
threshold singularities as the magnetic field is increased.
In Fig. \ref{fig:chinn} we 
plot $\chi_{nn}(\omega,2k_{F,\sigma}+q)$ for $\beta_c=\beta_s=1$ and
$\alpha=0.851$, which corresponds to the half-filled Hubbard model in
zero magnetic field at $U=1$ ($v_c=2.15 ta_0$, $v_s=1.83ta_0$).
At $\omega\sim 8M$ one can just see that the threshold is below the
curve $\sqrt{4M^2 +v_c^2 q^2}$.

\section{Excitons: $\beta_c^2<\frac{1}{2}$}
\label{sec:excitons}
In the regime $\beta_c^2<\frac{1}{2}$ soliton and antisoliton can form
excitonic bound states which for the SGM are known as breathers. There
are
\begin{equation}
{\cal N}= \left[ \frac{1-\beta^2}{\beta^2}\right]
\label{Nex}
\end{equation}
different types of excitons, where $[ x ]$ denotes the integer part
of~$x$. We denote the different excitons by $e_1,e_2,\ldots$.
The exciton gaps are given by 
\begin{equation}
M_n=2M \sin(n\pi\xi/2)\ ,\quad n=1,\ldots ,{\cal N}.
\label{exgap}
\end{equation}
It follows from the transformation properties under $\Phi_c\to-\Phi_c$
that only the ``odd'' excitons $e_{2n+1}$ couple to the density
operator. For simplicity we take $\beta_c>1/2$ from now, in which case
we only have to consider the first exciton $e_1$. In this regime of
$\beta_c$ the leading contributions of the spectral sum to the
dynamical density-density correlator are given by
\bea
\chi_{nn}(\omega,k)=\chi_{nn}^{\rm
exc}(\omega,k)+\chi_{nn}^{s\bar{s}}(\omega,k)\ ,
\label{chi.tot}
\eea
where $\chi_{nn}^{\rm exc}(\omega,k)$ and
$\chi_{nn}^{s\bar{s}}(\omega,k)$ denote the contributions from
intermediate states with one exciton and many spinons and one soliton,
one antisoliton and may spinons respectively. We have already
calculated $\chi_{nn}^{s\bar{s}}(\omega,k)$ above. For small $k$ it is
given by \r{q0} and for $k\approx k_{F,\sigma}$ by \r{differentv}.
The exciton contribution can be calculated by the same method
\cite{FF} and we now present the results.

\subsection{Behavior around $k=0$}

\begin{figure}[ht]
\begin{center}
\epsfxsize=0.45\textwidth
\epsfbox{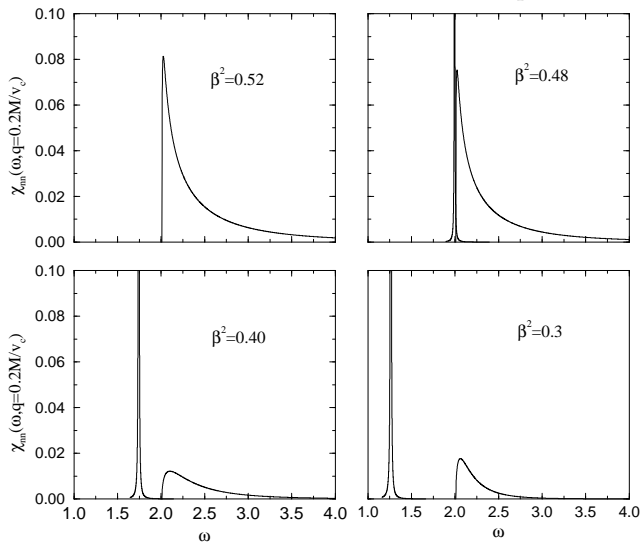}
\end{center}
\caption{\label{fig:q0}
$\chi_{nn}(\omega,q=0.2M/v_c)$ for  different values of $
\beta_c$. We have broadened the delta
function by convoluting it with a Lorentzian in order to exhibit the
transfer of spectral weight from the soliton-antisoliton continuum to
the coherent breather peak.
}
\end{figure}

Here the exciton is visible as a sharp $\delta$-function peak at an
energy below the soliton-antisoliton scattering continuum
\bea
\chi_{nn}^{\rm exc}(\omega,q)=A^2 g_0 \frac{v_c^2q^2}{\omega}
\delta(\omega-\sqrt{v_c^2q^2+M_1^2})\ ,
\label{excq=0}
\eea
where
\bea
g_0&=&\frac{\pi\lambda^2\xi^2}{\sin^2(\pi\xi)}\ ,\nn
\lambda&=&2 \cos\left(\frac{\pi\xi}{2}\right)\sqrt{2\sin
\left(\frac{\pi\xi}{2}\right)}
\exp\left(-\int_0^{\pi\xi}\frac{dt}{2\pi}\frac{t}{\sin t}\right).\nn
\eea
The result \r{excq=0} is again related by the equations of motion
\r{eom} to the corresponding contribution to the optical conductivity
\cite{EGJ}. The dynamical density susceptibility, (\ref{chi.tot}), 
for $v_cq=0.2M$, 
is plotted in Fig.~\ref{fig:q0}. For $\beta_c < 1/\sqrt{2}$ the breather 
contribution (\ref{excq=0}) appears and  the spectral weight is gradually 
transferred from the soliton-antisoliton continuum to the coherent peak.

\subsection{Behavior around $q=2k_{F,\sigma}$}
\begin{figure}[ht]
\begin{center}
\epsfxsize=0.45\textwidth
\epsfbox{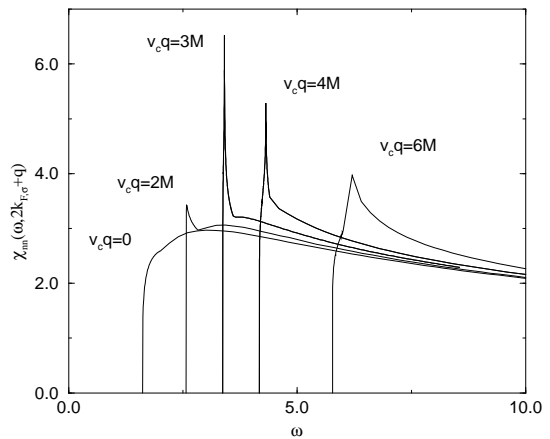}
\end{center}
\caption{\label{fig:exc.2kf}
$\chi_{nn}(\omega,2k_{F,\sigma}+q)$ for $\alpha=0.8$, $\beta_s=1$ and 
$\xi=0.6$. For $\omega\approx \omega_{\rm cusp}$ one can clearly see
the cusp due to the breather contribution. As $v_cq$ approaches
$v_cQ^\prime\approx 2.157M$ the cusp turns into a singularity.
}
\end{figure}

In the vicinity of $q=2k_{F,\sigma}$ the exciton contributes to
the dynamical density-density correlation function via an
exciton-spinon scattering continuum with threshold
\bea
E_{\rm thres}&=&\min_{q}\left[
v_s|P-q|+\sqrt{M_1^2+v_c^2q^2}\right]\nn
&=&\cases{
\sqrt{M_1^2+v_c^2 P^2} & {\rm if}$\ |P|\leq Q^\prime$\cr
v_s|P|+M_1\sqrt{1-\alpha^2} & {\rm if}$\ |P|\geq Q^\prime$\cr
}\ ,
\label{thres2}
\eea
where
\be
Q^\prime = \frac{M_1 v_s}{v_c\sqrt{v_c^2 - v_s^2}}\ .
\ee
The exciton contribution to the dynamical density~-density correlation
function is given by
\bea
&&\chi_{nn}^{\rm exc}(\omega,2k_{F,\sigma}+q)\approx
\frac{2\pi Z_1{A^\prime}^2}{\Gamma^2(d)(2v_s)^{2d-1}}\nn
&&\times
\left[\frac{\lambda \xi }{2\cos(\pi\xi/2)}\right]^2
 \int_{-\infty}^\infty d\theta
\left(\Sigma\Sigma^\prime\right)^{d-1}
\Theta(\Sigma)\ \Theta(\Sigma^\prime),\nn
\label{differentvex}
\eea
where
\bea
\Sigma&=&\omega-v_sq-M_1\left[\cosh(\theta)-\alpha\sinh(\theta)\right],\nn
\Sigma^\prime&=&\omega+v_sq-M_1
\left[\cosh(\theta)+\alpha\sinh(\theta)\right].
\label{oop}
\eea
One readily deduces by inspection of
equations \r{oop} and \r{differentvex} that for fixed $q$ the exciton
contribution to $\chi_{\nn}$ exhibits a {\sl cusp} at a frequency
\be
\omega_{\rm cusp}=\sqrt{M_1^2+v_c^2q^2}\ .
\ee
If we fix the momentum transfer to be $q=Q^\prime$, this cusp turns
into a singularity. In order to exhibit these interesting features we plot
$\chi_{nn}(\omega,2k_{F,\sigma}+q)$ for $\xi=0.6$, $\alpha=0.8$ 
and different values of $q$ in Fig.\ref{fig:exc.2kf}.

\section{Summary}
\label{sec:summ}
We have studied the dynamical density-density response of half-filled
1D Mott insulators for the case where the Mott gap is small compared
to the hopping matrix element $t$. We have allowed for the presence of
a magnetic field that partially magnetizes the ground state.
Unlike the spectral function, the density-density response function
does not exhibit prominent, dispersing features associated with the
spin and charge degrees of freedom respectively. Due to the momentum
dependence of the matrix elements in the gapped charge sector,
$\chi_{nn}(\omega,k)$ tends to zero as the threshold is approached from
above: irrespective of the magnitude of the applied magnetic field
there are no threshold singularities. 

In the parameter region of the extended Hubbard model where excitons
are formed, the dynamical density-density response exhibits a {\sl
cusp} at some specific value of energy transfer for momentum transfers
$k$ close to $2k_{F,\sigma}$. This cusp turns into a singularity for one
particular value of $k$. These features should be experimentally
observable in small-gap quasi-1D Mott insulators.

\acknowledgments
FHLE ackowledges support by the DOE (DE-AC02-98 CH 10886) and the EPSRC
(grants AF/100201 and GR/N19359). We thank M. Fabrizio, F. Gebhard and
A.A. Nersesyan for discussions.


\begin{thebibliography}{99}
\bibitem{conteq}
The contribution of \r{nu} to $\chi_{nn}$ is easily calculated by the
formfactor method and is found to be finite for $q\to 0$ at fixed
$\omega$ (this is obvious as \r{nu} is a Lorentz scalar and the
contribution to the correlation function therefore is a function of
$s^2=\omega^2-v^2q^2$). This would imply an infinite optical
conductvity by \r{eom}, in contradiction to Ref.\onlinecite{cet}.
\end{thebibliography}
\end{document}